\documentclass[aps,twocolumn,superscriptaddress,longbibliography]{revtex4-1}

\usepackage{hyperref}
\hypersetup{colorlinks,allcolors=blue}
\usepackage{bm}
\usepackage{amsmath}
\usepackage{graphicx} 
\usepackage{color}
\usepackage{siunitx} 
\usepackage[all]{hypcap} 

\newcommand{\jone}{$j_\mathrm{eff} = 1/2$}

\begin{document}

\title{A \texorpdfstring{\emph{j}\textsubscript{eff} = 1/2}{\emph{j}\textsubscript{eff} = 1/2} Pseudospinon Continuum in CaIrO\texorpdfstring{\textsubscript{3}}{3}}

\author{Matteo~Rossi}
\email{rossim@stanford.edu}
\altaffiliation[Present address: ]{Stanford Institute for Materials and Energy Sciences, SLAC National Accelerator Laboratory, 2575 Sand Hill Road, Menlo Park, CA 94025, USA.}
\affiliation{ESRF -- The European Synchrotron, 71 Avenue des Martyrs, CS 40220, F-38043 Grenoble, France}
\affiliation{Dipartimento di Fisica, Politecnico di Milano, Piazza Leonardo da Vinci 32, I-20133 Milano, Italy}

\author{Pietro~Marabotti}
\affiliation{Dipartimento di Fisica, Politecnico di Milano, Piazza Leonardo da Vinci 32, I-20133 Milano, Italy}

\author{Yasuyuki~Hirata}
\altaffiliation[Present address: ]{Department of Applied Physics, National Defense Academy of Japan, 1-10-20 Hashirimizu, Yokosuka, Kanagawa 239–8686, Japan.}
\affiliation{Institute for Solid State Physics, The University of Tokyo, Kashiwa, Chiba 277-8581, Japan}

\author{Giulio~Monaco}
\affiliation{Dipartimento di Fisica, Universit\`{a} di Trento, via Sommarive 14, I-38123 Povo (TN), Italy}

\author{Michael~Krisch}
\affiliation{ESRF -- The European Synchrotron, 71 Avenue des Martyrs, CS 40220, F-38043 Grenoble, France}

\author{Kenya~Ohgushi}
\altaffiliation[Present address: ]{Department of Physics, Tohoku University, 6-3 Aramaki-Aoba, Aoba-ku, Sendai, Miyagi 980-8578, Japan.}
\affiliation{Institute for Solid State Physics, The University of Tokyo, Kashiwa, Chiba 277-8581, Japan}

\author{Krzysztof~Wohlfeld}
\affiliation{Institute of Theoretical Physics, Faculty of Physics, University of Warsaw, Pasteura 5, PL-02093 Warsaw, Poland}

\author{Jeroen~{van~den~Brink}}
\affiliation{Institute for Theoretical Solid State Physics, IFW Dresden, Helmholtzstrasse 20, D-01069 Dresden, Germany}
\affiliation{Department of Physics, Technical University Dresden, D-01062 Dresden, Germany}

\author{Marco~{Moretti~Sala}}
\email{marco.moretti@polimi.it}
\affiliation{ESRF -- The European Synchrotron, 71 Avenue des Martyrs, CS 40220, F-38043 Grenoble, France}
\affiliation{Dipartimento di Fisica, Politecnico di Milano, Piazza Leonardo da Vinci 32, I-20133 Milano, Italy}

\begin{abstract}
In so-called \jone~systems, including some iridates and ruthenates, the coherent superposition of $t_\mathrm{2g}$ orbitals in the ground state gives rise to hopping processes that strongly depend on the bond geometry. Resonant inelastic x-ray scattering (RIXS) measurements on CaIrO$_3$ reveal a prototypical \jone~\emph{pseudo}spinon continuum, a hallmark of one-dimensional (1D) magnetic systems despite its three-dimensional crystal structure. The experimental spectra compare very well to the calculated magnetic dynamical structure factor of weakly coupled spin-1/2 chains. We attribute the onset of such quasi-1D magnetism to the fundamental difference in the magnetic interactions between the \jone~\emph{pseudo}spins along the corner- and edge-sharing bonds in CaIrO$_3$.

\end{abstract}

\maketitle

\section{Introduction}

Spin-orbit-induced Mott insulators, including iridium oxides (iridates) and RuCl$_3$, host a large variety of physical properties, whose microscopic origin is intrinsic to the nature of the \jone~ground state and of the interactions it gives rise to \cite{Kim2008,Kim2009,WitczakKrempa2014,Rau2016}. Electronic interactions in insulating systems originate from virtual hopping processes, possibly through the ligands and, depending on the bond geometry, multiple hopping paths may exist. This possibility is enhanced in iridates, where the \jone~electronic wave function arises from a coherent superposition of all $t_\mathrm{2g}$ orbitals with mixed phases and spin components. As a consequence, magnetic interactions in corner-, edge-, or face-sharing octahedral iridates are drastically different \cite{Jackeli2009,Matsuura2014,Kugel2015,Khomskii2016,Takagi2019}.

Corner-sharing layered-perovskite iridates bear isotropic Heisenberg-like antiferromagnetic (AFM) exchange interactions between \jone~\emph{pseudo}spins, with a small symmetric anisotropic contribution \cite{Jackeli2009}. Indeed, materials featuring corner-sharing IrO$_6$ octahedra on a quasi-two-dimensional (2D) square lattice, including Sr$_2$IrO$_4$ \cite{Crawford1994,Cao1998,Kim2009,Boseggia2013}, Sr$_3$Ir$_2$O$_7$ \cite{Cao2002,Kim2012a,Boseggia2012,Boseggia2012a} and Ba$_2$IrO$_4$ \cite{Boseggia2013a,MorettiSala2014d}, display magnetic properties reminiscent of cuprates \cite{Kim2012,Kim2014a,Kim2015,DeLaTorre2015,Yan2015,Gretarsson2016,Liu2016,Jeong2017,Pincini2017} and attracted the interest of the scientific community -- even more so when unconventional superconductivity was predicted in such iridates \cite{Wang2011,Watanabe2013,Yang2014,Meng2014}. Distortions of the ideal bond geometry affect both the nature and strength of the exchange interactions. For example, if the corner-sharing bond lacks inversion symmetry, an antisymmetric exchange anisotropy is allowed; Dzyaloshinski-Moriya interaction sets in and induces a finite canting of the magnetic moments away from the perfectly collinear structure, thus giving rise to weak ferromagnetism in Sr$_2$IrO$_4$ \cite{Kim2009,Boseggia2013}.

For edge-sharing IrO$_6$ octahedra, instead, the hopping paths via the two bridging oxygens interfere in a destructive manner and can cause the cancellation of the isotropic exchange interaction \cite{Jackeli2009}. As a result, a weak Ising-like anisotropic interaction dominates, which gives rise to Kitaev-like spin-spin exchange \cite{Chun2015,Hermanns2018,Takagi2019}. Particularly interesting are materials composed of edge-sharing IrO$_6$ octahedra on a honeycomb (Na$_2$IrO$_3$ \cite{Singh2010,Singh2012,Chun2015,Mehlawat2017}, $\alpha$-Li$_2$IrO$_3$ \cite{Singh2012}), hyper- ($\beta$-Li$_2$IrO$_3$ \cite{Takayama2015}) and stripy- ($\gamma$-Li$_2$IrO$_3$ \cite{Modic2014}) honeycomb structure as they sustain Kitaev frustration and are therefore candidates for unconventional quantum ground states and excitations such as spin liquid ground states that can harbour Majorana fermions \cite{Kitaev2006,Hermanns2018,Takagi2019}. However, the Kitaev interaction is usually not very strong and other magnetic couplings, otherwise irrelevant, can induce long-range magnetically-ordered phases \cite{Chaloupka2010,Choi2012,Chaloupka2013,Banerjee2016,Winter2016}. Face-sharing iridates, on the other hand, are not yet fully explored, but allow for intriguing properties. For example, iridates featuring Ir$_2$O$_9$ bioctahedra on a triangular lattice, including Ba$_3$(Ti,Zr,Ce,Pr,Tb)Ir$_2$O$_9$ \cite{Doi2004,Sakamoto2006}, show excessively small values of the effective magnetic moments possibly due to covalency effects \cite{Kugel2015,Khomskii2016,Revelli2019}. 

\begin{figure}
	\centering
	\includegraphics[width=0.9\columnwidth]{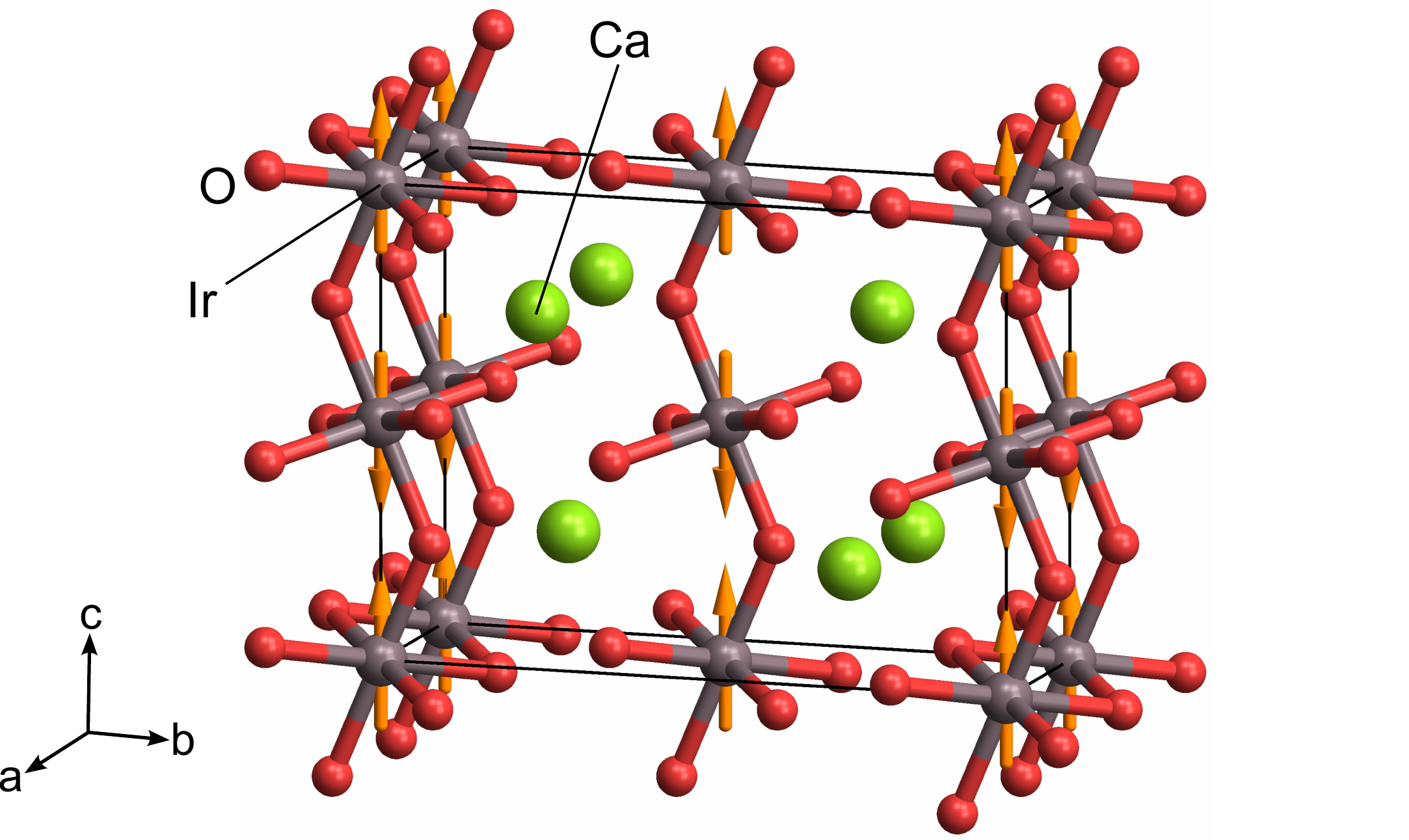}
	\caption{\label{fig:crystal} Crystal and magnetic structure of CaIrO$_3$. Ir ions (grey spheres) are surrounded by octahedra of O ions (red spheres) sharing an edge and a corner along the $a$ and $c$ axes, respectively. Layers of Ca ions (green spheres) separate the $ac$ planes. \jone~pseudospins are displayed as orange arrows.}
\end{figure}

The various types of interactions on the bond geometries described above are characterized by very distinct energy scales: the Kitaev interaction in edge-sharing systems is about an order of magnitude smaller than the Heisenberg exchange in corner-sharing systems. However, material-dependent properties could also affect the origin and magnitude of the superexchange, as discussed above. Thus, we here consider the post-perovskite CaIrO$_3$, whose crystal structure features both edge-sharing IrO$_6$ octahedra along the $a$ axis and corner-sharing octahedra along the orthogonal $c$ direction (see Fig.~\ref{fig:crystal}). Even if the crystal structure is quasi 2D, with IrO planes separated by Ca spacers, we find evidence of quasi 1D magnetic dynamics. By means of resonant inelastic x-ray scattering (RIXS), we observe the dispersion of \emph{pseudo}spinons in a \jone~spin-orbit Mott insulator for the first time. The reduced dimensionality is strictly related to the nature of the \jone~ground state on the bond geometries and not guaranteed by crystallography alone. Our results are in agreement with the proposed striped AFM order \cite{Ohgushi2013} and theoretical predictions by \emph{ab initio} quantum chemical calculations \cite{Bogdanov2012}. Our findings provide a new direction to tailor the magnetic properties of \jone~systems while pertaining a higher-dimensional lattice.

\section{Experimental details}

Single crystals of CaIrO$_3$ were grown using the flux method described in Ref.~\onlinecite{Ohgushi2013}. Crystals have a needle shape with length of $\sim 1$ mm and diameter of $\sim 0.1$ mm. CaIrO$_3$ grows in the post-pervoskite crystal structure (space group $Cmcm$), with lattice parameters $a = 3.147$ \AA, $b = 9.863$ \AA~and $c = 7.299$ \AA~\cite{Ohgushi2006,Cheng2011}. IrO$_6$ octahedra share an edge along the $a$ direction and a corner along the $c$ direction (Fig.~\ref{fig:crystal}). Below the N\'{e}el temperature $T_N \approx 110$ K \cite{Ohgushi2006,Cheng2011,Gunasekera2015,Gunasekera2015a}, \jone~pseudospins order in a striped magnetic structure: magnetic moments are mostly aligned parallel to the $c$ axis and are ferromagnetically (antiferromagnetically) coupled along the $a$ ($c$) direction \cite{Ohgushi2013}, as shown in Fig.~\ref{fig:crystal}. 

Iridium $L_3$ edge ($\sim 11.22$ keV) RIXS measurements were performed at the inelastic X-ray scattering beamline ID20 of ESRF -- The European Synchrotron (Grenoble, France) \cite{MorettiSala2013,MorettiSala2018}. The incident beam was monochromated using a Si(844) back-scattering channel-cut crystal. The spectrometer is based on the Rowland circle geometry and is equipped with a single $R=1$ m Si(844) diced crystal analyzer and a two-dimensional Maxipix detector \cite{Ponchut2011}. A mask with circular aperture of 30 mm was installed in front of the analyzer to improve the momentum resolution. The overall energy and momentum resolutions were 35 meV (full width at half maximum) and approximately 0.2 \AA$^{-1}$, respectively. The sample was cooled by means of a cryostat using He as exchange gas \cite{vanderLinden2016}.

\section{Results and discussion}

The RIXS response of CaIrO$_3$ has been discussed by Moretti Sala \emph{et al.} \cite{MorettiSala2014b}, who investigated the excitations at energy losses above 0.4 eV and ascribed those to electronic transitions within the $t_\mathrm{2g}$ manifold. In agreement with quantum chemical calculations \cite{Bogdanov2012,KimSW2015,Singh2016}, the analysis of RIXS data based on a single-ion model \cite{Kim2008,Ament2011,MorettiSala2014a} shows that the ground state of CaIrO$_3$ departs from the ideal \jone~wave function, with the single hole unequally distributed among the $t_\mathrm{2g}$ orbitals \cite{MorettiSala2014b}. In particular, it was found that the hole has predominant $(|d_{yz},\pm\rangle \mp i|d_{xz},\pm\rangle)/\sqrt{2}$ character \cite{MorettiSala2014b}. A priori, this might have implications on the bond-geometry dependence of the magnetic interactions; however, we note that the exchange processes in CaIrO$_3$ mostly involve the Ir $d_{yz}$ and $d_{xz}$ orbitals in both bond geometries and, since the phase and relative occupancies of these states are preserved, the nature of the dominant exchange interactions should be unaffected. We note that the hopping involving the $e_\mathrm{g}$ states can be safely neglected due to the large cubic crystal field (typically $\sim 3$ eV in iridates \cite{MorettiSala2014c}).

\begin{figure*}[ht]
	\centering
	\includegraphics[width=\textwidth]{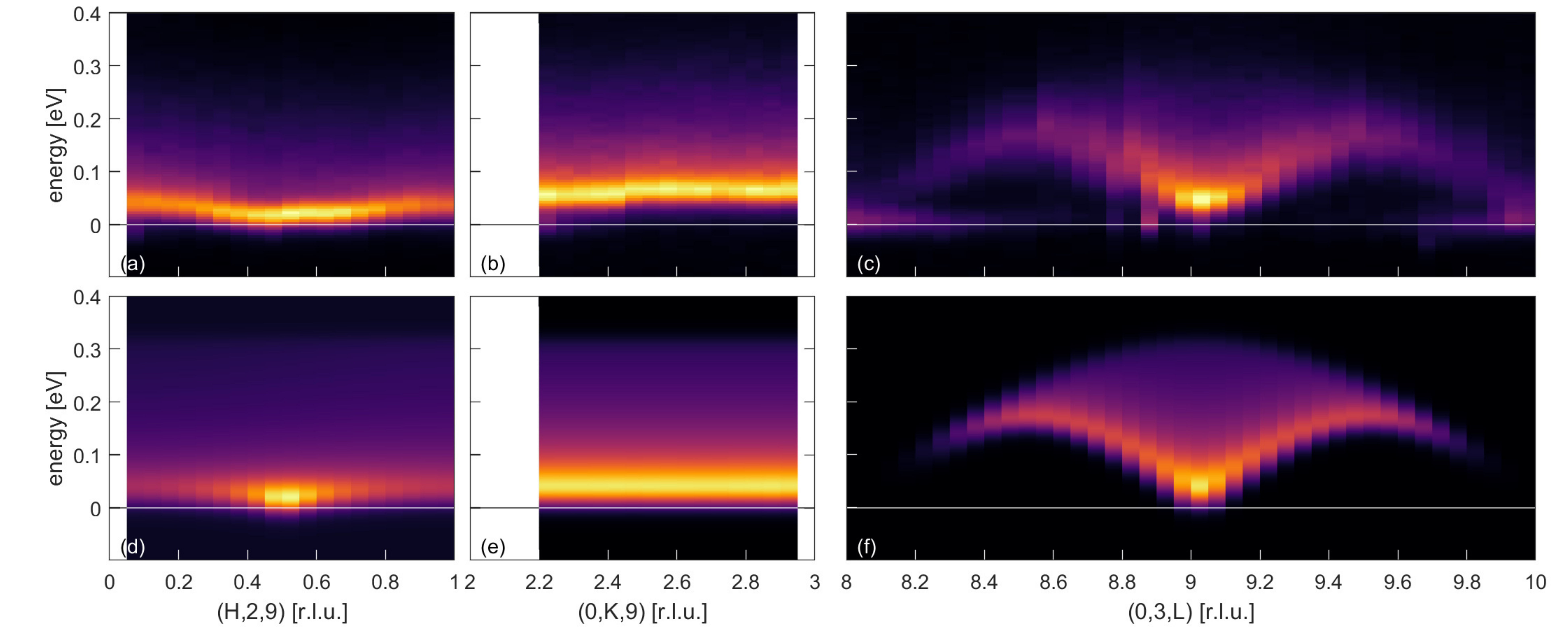}
	\caption{\label{fig:RIXS_maps} RIXS intensity maps of CaIrO$_3$ measured at 40 K as a function of momentum transfer along the $H$ (a), $K$ (b), and $L$ (c) high-symmetry directions of the reciprocal lattice. Panels (d), (e) and (f) display the magnetic dynamical structure factor of weakly coupled $S = 1/2$ chains computed for $J_c = 98$ meV and $J_a = -4.4$ meV.}
\end{figure*}

We here focus on the spectral features at energy losses below 0.4 eV and on their momentum dependence. RIXS measurements along high-symmetry directions in reciprocal space were taken at 40 K and are shown in Fig.~\ref{fig:RIXS_maps}(a), (b) and (c). We first note that an excitation continuum extending up to $\sim 0.3$ eV appears in all high-symmetry directions and that it looks ``gapped'' everywhere in reciprocal space. The continuum behaves very differently in the three orthogonal directions. Specifically, along the $H$ (edge-sharing) [Fig.~\ref{fig:RIXS_maps}(a)] and $K$ [Fig.~\ref{fig:RIXS_maps}(b)] directions, the energy of the main peak is minimum and shows little and no momentum dependence, respectively. On the opposite, along the $L$ (corner-sharing) direction [Fig.~\ref{fig:RIXS_maps}(c)], the RIXS spectra show a pronounced momentum dependence and the dispersion of the continuum is characterized by a double-arched shape: the low-energy boundary has a periodicity of 1 r.l.u., while the high energy envelope has double periodicity. We note immediately that the magnetic excitations of CaIrO$_3$ are markedly distinct from the ones of other layered iridates such as Sr$_2$IrO$_4$ \cite{Kim2012}, Sr$_3$Ir$_2$O$_7$ \cite{Kim2012a,MorettiSala2015} and Na$_2$IrO$_3$ \cite{Chun2015}. Instead, the momentum dependence of RIXS spectra observed in Fig.~\ref{fig:RIXS_maps}(c) is reminiscent of the dynamics of 1D AFM $S = 1/2$ chains \cite{Giamarchi2004,Karbach1997,Zaliznyak2004,Lake2005,Caux2006,Lake2009,Schlappa2012,Mourigal2013,Wu2016,Bera2017,Fumagalli2020}. In these systems, magnetic excitations are quasiparticles called spinons \cite{Giamarchi2004}: in a simple picture, spinons are domain walls that separate two degenerate regions of 1D AFM chains with out-of-phase spin order; they are created in pairs when a spin is excited within the chain. Since the two spinons can freely propagate along the chain with no energy cost, the $S = 1$ excitation dissociates into two independent entities carrying $S = 1/2$ each. This process, called fractionalization, is a hallmark of 1D AFM $S = 1/2$ chains. In the non-interacting picture, a two-spinon excitation can be considered as the excitation of two, independent single spinons, giving rise to the peculiar energy-momentum continuum \cite{Giamarchi2004}. Based on the similarity between the two-spinon continuum and the excitation spectrum of CaIrO$_3$, we assign the latter to magnetic excitations of 1D AFM chains. In the case of CaIrO$_3$, precautions should be taken because the magnetic moments are not $S=1/2$ spins, but pseudospins that carry orbital angular momentum apart from the spin one. However, Fig.~\ref{fig:RIXS_maps}(c) is an unmistakable signature of the quasi-1D nature of the magnetic excitations of CaIrO$_3$ and, as shown below, provides the first experimental evidence of fractionalization of \emph{pseudo}spinons in a spin-orbit Mott insulator. CaIrO$_3$ is also one of the very few compounds exhibiting fractionalization that are not based on $3d$ transition metals.

In order to verify the above hypothesis and quantify the strength of the magnetic couplings in CaIrO$_3$, we compare our RIXS results to the calculated magnetic dynamical structure factor $S(\mathbf{Q},\omega)$. We first consider the magnetic dynamical structure factor $S_\mathrm{1D}(q_z,\omega)$ of a 1D AFM $S=1/2$ chain with isotropic magnetic coupling $J_c>0$, where $q_z = 2\pi L/c$ is the momentum transfer along the chain direction: it was derived by M\"uller \emph{et al.}~\cite{Mueller1981} and is characterized by a gapless double-arched excitation continuum, without momentum dependence perpendicular to the chain. However, the weak dispersion along the $H$ direction shown in Fig. \ref{fig:RIXS_maps}(a) suggests that a minimal model aiming at describing CaIrO$_3$ should include a magnetic coupling perpendicular to the chain direction $J_a$, which we assume to be isotropic for simplicity and negative to account for the FM alignment of the magnetic moments along the $a$ direction \cite{Ohgushi2013}. Within the random phase approximation \cite{Bockquet2001}, the relation between the magnetic susceptibility of weakly coupled chains $\chi(\mathbf{Q},\omega)$ and that of a 1D $S=1/2$ chain $\chi_\mathrm{1D}(q_z,\omega)$ is \cite{Scalapino1975,Schulz1996,Bockquet2001}:
\begin{equation}
\chi(\mathbf{Q},\omega) = \frac{\chi_\mathrm{1D}(q_z,\omega)}{1 - 2J_a \cos(q_x) \chi_\mathrm{1D}(q_z,\omega)},
\end{equation}
where $q_x=2\pi H/a$ is the momentum transfer perpendicular to the chain direction. The magnetic dynamical structure factor is related to the imaginary part of the magnetic susceptibility: $S(\mathbf{Q},\omega) = -\chi''(\mathbf{Q},\omega)$. A similar relationship holds for $S_\mathrm{1D}(q_z,\omega)$ and $\chi_\mathrm{1D}''(q_z,\omega)$. This approach has been extensively used to simulate the magnetic dynamical structure factor of systems characterized by weakly coupled magnetic chains as measured by inelastic neutron scattering \cite{Zheludev2003,Enderle2010,Leiner2018}. 

Before any comparison to the experimental results is made, the calculated $S(\mathbf{Q},\omega)$ should be broadened by the finite energy and momentum resolutions. The latter is determined by the analyzer surface and depends on the scattering geometry, as detailed in Ref.~\onlinecite{MorettiSala2018}. The effect of a finite momentum resolution is critical near the Brillouin zone center ($q_z = 0$) and close to the zone boundaries ($q_z = \pm \pi$), where the dispersion of magnetic excitations is the steepest: indeed, it seems to artificially open a gap in the magnetic excitations spectrum, which, for a system of $S=1/2$ spins interacting through isotropic couplings only, is gapless \cite{Karbach1997,Caux2006}. 

\begin{figure}
	\centering
	\includegraphics[width=\columnwidth]{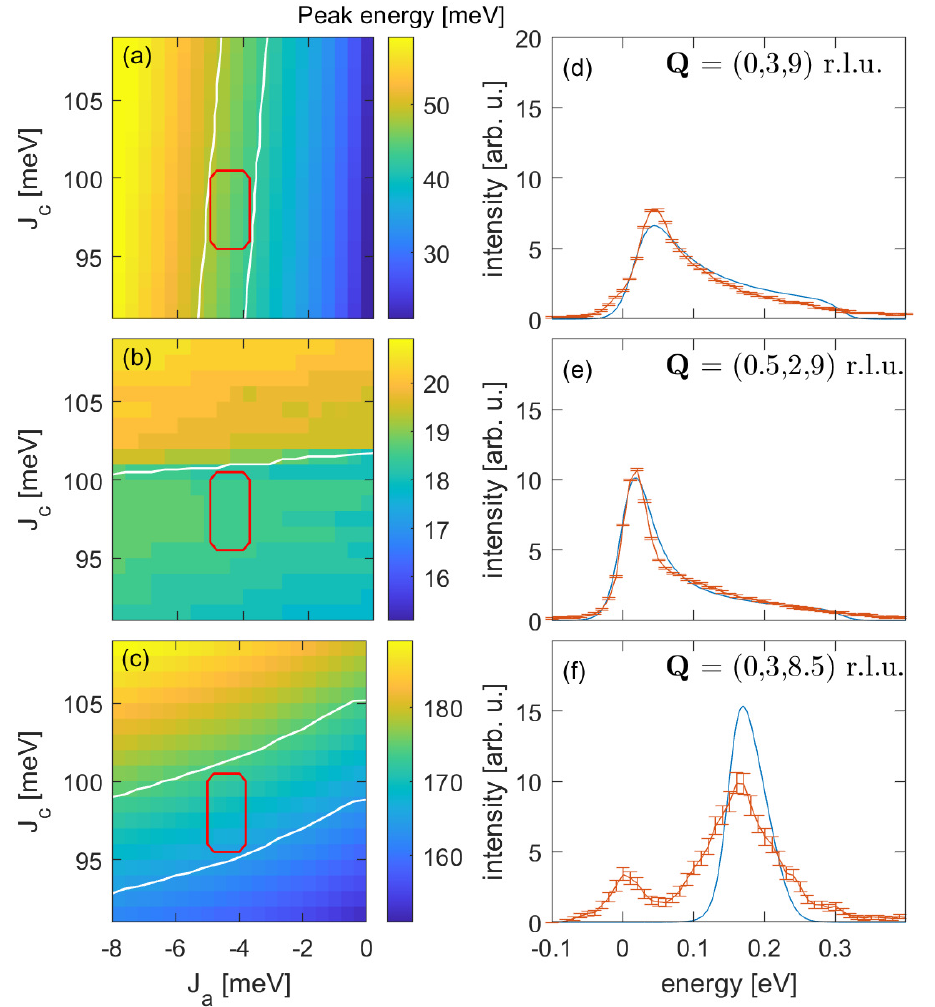}
	\caption{\label{fig:parameters} Calculations of the magnetic peak position from $S(\mathbf{Q}, \omega)$ as a function of $J_a$ and $J_c$ for (a) $\mathbf{Q} = (0, 3, 9)$ r.l.u., (b) $\mathbf{Q} = (0.5, 2, 9)$ r.l.u. and (c) $\mathbf{Q} = (0, 3, 8.5)$ r.l.u.. The experimental peak position falls within the domains represented by white lines. The intersection of these three domains is enclosed by red circles. (d), (e) and (f) show the comparison between RIXS spectra (red) and $S(\mathbf{Q}, \omega)$ (blue solid lines) calculated for the optimal values $J_c = 98$ meV and $J_a = -4.4$ meV.}
\end{figure}

In Fig.~\ref{fig:parameters}(a), (b) and (c), we extracted the energies of the magnetic peak from the calculated $S(\mathbf{Q},\omega)$ as a function of $J_a$ and $J_c$ for three representative transferred momenta; white contour lines correspond to the magnetic peak position as determined by RIXS, \emph{i.e.} $45 \pm 3$ meV at $\mathbf{Q} = (0, 3, 9)$ r.l.u., $18 \pm 1$ meV at $\mathbf{Q} = (0.5, 2, 9)$ r.l.u. and $170 \pm 5$ meV at $\mathbf{Q} = (0, 3, 8.5)$ r.l.u.. The intersection of these domains is a very small region of the \{$J_a$, $J_c$\} parameter space enclosed by the red solid line. In particular, the best fit to the data is obtained for $J_a=-4.4$ meV and $J_c=98$ meV. The magnitudes of the magnetic couplings beautifully confirm the predictions of quantum chemical calculations \cite{Bogdanov2012}. The magnetic dynamical structure factors capture the corresponding RIXS spectra, as shown in Fig.~\ref{fig:parameters}(d), (e) and (f). We note that the magnetic excitations measured at $\mathbf{Q} = (0,3,8.5)$ r.l.u. are broader than the corresponding dynamical structure factor (Fig.~\ref{fig:parameters}(f)). This is surprising given the excellent agreement at other momenta and will be the subject of further investigation. The full momentum dependence of the calculated $S(\mathbf{Q},\omega)$ along the orthogonal $H$, $K$ and $L$ directions in reciprocal space is compared to the RIXS spectra in Fig.~\ref{fig:RIXS_maps}: the magnetic dynamical structure factor of weakly coupled ($|J_a| \ll J_c$) 1D AFM $S=1/2$ chains satisfactorily reproduces the main features observed in our experimental results.

\begin{figure}
	\centering
	\includegraphics[width=\columnwidth]{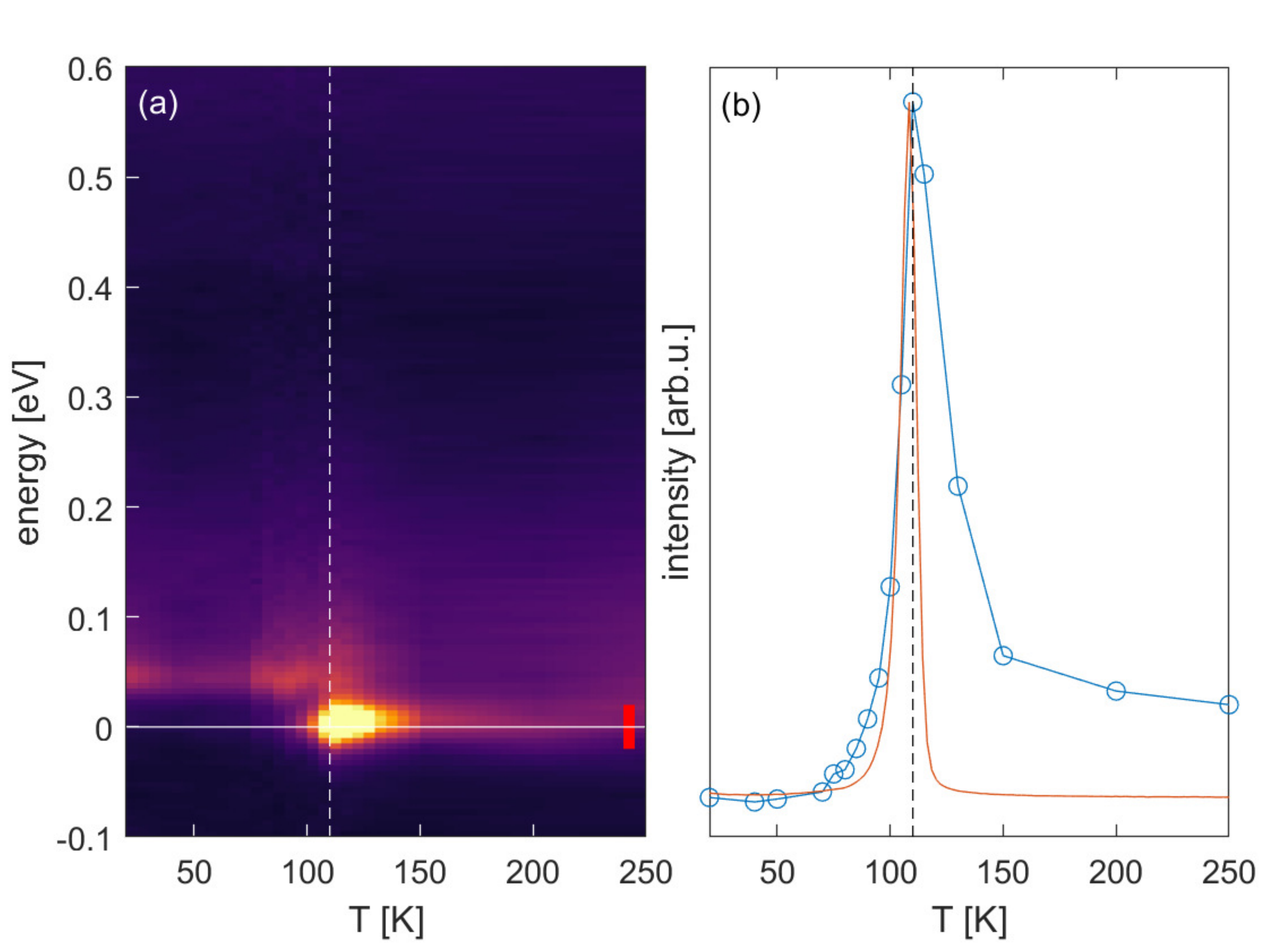}
	\caption{\label{fig:T_dependence} (a) RIXS intensity map as a function of temperature at $\mathbf{Q} = (1,2,9)$ r.l.u. showing the strong maximum of the quasielastic signal at the N\'{e}el temperature (dashed line). (b) Temperature dependence of the quasielastic region of the RIXS spectra (blue open circles) obtained by integrating the RIXS signal in the energy window of $\pm 20$ meV, represented by the red vertical bar in panel (a). The red solid line is the imaginary part of the quasi-static magnetic susceptibility, reported from Refs.~\onlinecite{Gunasekera2015,Gunasekera2015a}.}
\end{figure}

Before concluding, we report in Figure~\ref{fig:T_dependence}(a) the temperature dependence of RIXS spectra at momentum transfer $\mathbf{Q} = (1, 2, 9)$ r.l.u.. The two-spinon continuum is present at all temperatures. At high temperature, the spectra are characterized by a barely visible elastic line. When the sample is cooled down, the elastic peak intensity increases with a maximum at the magnetic phase transition and then decreases to complete suppression below 70 K. The trend of the elastic line is captured by the blue circles plotted in Fig.~\ref{fig:T_dependence}(b), which reports the temperature dependence of the RIXS signal integrated over an energy window of $\pm 20$ meV centered around zero energy loss, highlighted as a red rectangle in Fig.~\ref{fig:T_dependence}(a). Its resemblance with the temperature dependence of the imaginary part of the quasi-static magnetic susceptibility (red solid line) \cite{Gunasekera2015,Gunasekera2015a} suggests that also the elastic peak contains contributions from magnetic fluctuations. These are characterized by very small magnetic couplings and their temperature evolution catches the transition from the paramagnetic to the (3D) long-range AFM ordered phase, although $\mathbf{Q} = (1,2,9)$ r.l.u. does not correspond to a magnetic Bragg peak.

For the sake of completeness, we address two main limitations of our approach: i) higher-order spinon excitations may give a minor contribution to the magnetic signal \cite{Caux2006}, but are neglected in our calculations for simplicity because the core-hole lifetime at the Ir $L_3$ edge is very small; ii) in the calculation of the magnetic dynamical structure factor, we considered two isotropic magnetic couplings only. However, the $C_\mathrm{2h}$ and $C_\mathrm{2v}$ point group symmetry of the edge- and corner-sharing bonds allows for symmetric \cite{Jackeli2009,Matsuura2014} and antisymmetric (Dzyaloshinskii–Moriya) \cite{Moriya1960} anisotropic interactions along the $a$ and $c$ directions, respectively. The latter is responsible for the canting of the magnetic moments observed in CaIrO$_3$ \cite{Ohgushi2013}, while the former possibly includes the Kitaev interaction. Unfortunately, a precise determination of all exchange parameters of CaIrO$_3$, especially very small ones, is not possible at the present stage. However, one can say that the anisotropic Kitaev interaction across the edge-sharing direction is much smaller than the Heisenberg interaction along the corner-sharing bond, which is consistent with Jackeli and Khaliullin's argument \cite{Jackeli2009}.

\section{Conclusions}

To summarize, we investigated the magnetic dynamics of CaIrO$_3$ by means of RIXS and found strong evidence for quasi-1D magnetic behavior. Specifically, we interpreted the momentum dependence of the magnetic dynamics as due to \jone~\emph{pseudo}spinon excitations. Theoretical calculations of the magnetic dynamical structure factor of weakly coupled AFM chains adequately reproduce the experimental observations. Our results provide the first evidence of a pseudospinon continuum in spin-orbit Mott insulators. Our findings are consistent with the prediction of a strong bond-dependence of the magnetic couplings in iridates \cite{Jackeli2009,Matsuura2014}, which lowers the effective dimensionality of magnetism in CaIrO$_3$ and provides a strategy to engineer the magnetic properties of \jone~materials within their higher-dimensional lattice.

\begin{acknowledgements}
M.~Rossi and M.~Moretti~Sala kindly acknowledge C.~Henriquet and R.~Verbeni for technical assistance during the experiment and D.~K.~Singh for providing the magnetic susceptibility data of CaIrO$_3$. M.~Rossi would like to thank S.~Toth for helpful discussions. K.~Wohlfeld acknowledges support of the National Science Center (NCN), Project No. 2016/22/E/ST3/00560. J.v.d.B acknowledges financial support from the German Research Foundation (Deutsche Forschungsgemeinschaft, DFG) via SFB1143 project A5 and supported by the DFG through the W\"{u}rzburg-Dresden Cluster of Excellence on Complexity and Topology in Quantum Matter - ct.qmat (EXC 2147, project id 39085490).

This is a pre-print of an article published in The European Physical Journal Plus. The final authenticated version is available online at: \url{https://doi.org/10.1140/epjp/s13360-020-00649-5}.
\end{acknowledgements}

\bibliography{biblio}

\end{document}